\documentclass[preprint,aps,superscriptaddress]{revtex4}
\usepackage{graphicx,amsfonts,amsmath}
\usepackage{dcolumn}
\usepackage{bm}

\renewcommand{\a}{\alpha}
\renewcommand{\b}{\beta}
\newcommand{\g}{\gamma}
\def\f{\phi}

\def\q{\theta}

\def\s{\sigma}

\newcommand{\pa}{\partial}                           
\newcommand{\hf}{\frac12}
\newcommand{\be}{\begin{equation}}
\newcommand{\ee}{\end{equation}}
\newcommand{\bea}{\begin{eqnarray}}
\newcommand{\eea}{\end{eqnarray}}

\begin{document}

\title{On the
finiteness of noncommutative supersymmetric QED$_3$  in the  covariant superfield formulation}

\author{A. F. Ferrari}
\author{H. O. Girotti}
\affiliation{Instituto de F\'{\i}sica, Universidade Federal do Rio Grande
do Sul, Caixa Postal 15051, 91501-970 - Porto Alegre, RS, Brazil}
\email{alysson, hgirotti, wen@if.ufrgs.br}

\author{M. Gomes}
\author {A. Yu. Petrov}
\altaffiliation[]{ Department of Theoretical Physics,
Tomsk State Pedagogical University
Tomsk 634041, Russia
(email: petrov@tspu.edu.ru)}
\affiliation{Instituto de F\'{\i}sica, Universidade de S\~{a}o Paulo,
 Caixa Postal 66318, 05315-970, S\~{a}o Paulo - SP, Brazil}
\email{mgomes, petrov, ajsilva@fma.if.usp.br}

\author{A. A. Ribeiro}
\affiliation{Instituto de F\'{\i}sica, Universidade Federal do Rio Grande
do Sul, Caixa Postal 15051, 91501-970 - Porto Alegre, RS, Brazil}
\author{A. J. da Silva}
\affiliation{Instituto de F\'{\i}sica, Universidade de S\~{a}o Paulo,
 Caixa Postal 66318, 05315-970, S\~{a}o Paulo - SP, Brazil}

\begin{abstract}
The three-dimensional noncommutative
supersymmetric QED is investigated within the superfield approach.
We prove the absence of UV/IR mixing in the theory at any
loop order and demonstrate its one-loop finiteness.
\end{abstract}

\maketitle
\newpage

During last years noncommutative gauge theories have been intensively
studied. The interest in this subject has deep motivations coming
mainly from string theory \cite{SW} (for a review see \cite{Nekr, Szabo}).
Different aspects of noncommutative gauge theories were discussed in
\cite{Mat,Hay,SJ,Bonora,FL,Gur,Armoni,Nichol}.

One of the most remarkable properties of noncommutative theories
consists of an unusual structure of divergences, the so called
UV/IR mixing, that could lead to the appearance of infrared divergences \cite{Minw,Mat}. It should be noticed that the  cancellation of quadratic and linear ultraviolet divergences in
commutative theories does not guarantee the absence of harmful infrared
divergences in their noncommutative counterparts \cite{Girotti1,WZ,Bichl,ourqed}.
The elimination of such divergences is crucial since they may obstruct the
development of a sound renormalization scheme, leading to the breakdown
of the perturbative series.

Based on experience, it is natural to expect that supersymmetry could improve this
situation \cite{Che,Mat}. In fact, the Wess-Zumino model \cite{WZ}
and the three-dimensional sigma-model \cite{sig} are renormalizable at
all loop orders.
This is furtherly supported by the results of \cite{Zanon} according
to which the one-loop effective action in ${\cal N}=1,2$ super-Yang-Mills
theory contains only logarithmic divergences while for ${\cal N}=4$ the theory  is one-loop finite
\cite{Zanon,Liu}.

In this paper we employ the covariant superfield formalism to study
noncommutative supersymmetric QED$_3$.
We will prove that this theory is free of nonintegrable UV/IR divergences
at any loop order. We shall also demonstrate that the model
is one-loop finite.

The action of the  three-dimensional ${\cal{N}}=1$ noncommutative supersymmetric
QED  is \cite{SGRS}
\begin{eqnarray}
S=\frac{1}{2g^2}\int d^5 z W^\alpha *W_\alpha\,, \label{2n}
\end{eqnarray}
where

\begin{eqnarray}
\label{sstr}
W_\beta =\frac{1}{2}D^\alpha D_\beta A_\alpha -
\frac{i}{2}[A^\alpha ,D_\alpha A_\beta ]-
\frac{1}{6}
[A^\alpha ,\{A_\alpha ,A_\beta \}]
\end{eqnarray}

\noindent
is a superfield strength constructed from the spinor
superpotential $A_\alpha $. Hereafter it is implicitly assumed that all
commutators and anticommutators are Moyal ones. In this work we consider only space-space noncommutativity, to evade unitarity problems \cite{gomis}.
This action is invariant under the gauge transformations

\bea
\label{gt}
\delta A_{\a}=D_{\a}K-i[A_{\a},K]\,.
\eea

\noindent
Then, we must add a gauge fixing term which we choose to be

\begin{eqnarray}
S_{GF}=-\frac{1}{4\xi g^2}\int d^5 z (D^\alpha A_\alpha )
D^2(D^\beta A_\beta )\,,
\end{eqnarray}

\noindent
leading to the  quadratic action

\begin{eqnarray}
\label{s2a}
S_2=\frac{1}{2g^2}\int d^5 z
\Big[\frac{1}{2}(1+\frac{1}\xi)A^\alpha
\Box A_\alpha
-\frac{1}{2}(1-\frac{1}\xi )A^\alpha i\pa_{\alpha\beta}D^2A^\beta \Big]\,.
\end{eqnarray}

The free gauge propagator is

\begin{eqnarray}
<A^\alpha (z_1)A^\beta (z_2)>=\frac{ig^2}2\left[
C^{\alpha\beta}\frac{1}{\Box}(\xi+1)-\frac{1}{\Box^2}
(\xi-1)i\pa^{\alpha\beta}D^2\right]\delta^5(z_1-z_2)\,,
\end{eqnarray}

\noindent
where $C^{\a\b}=-C_{\a\b}$ is the second-rank antisymmetric symbol
defined with the normalization $C^{12}=i$.
The most convenient  choice for the gauge fixing parameter  is
$\xi=1$, the Feynman gauge, in  which the propagator collapses to

\begin{eqnarray}
\label{lgauge}
<A^\alpha (z_1)A^\beta (z_2)>=ig^2C^{\alpha\beta}\frac{1}{\Box}
\delta^5(z_1-z_2)\,.
\end{eqnarray}

\noindent
The interaction part of the classical action in the pure gauge sector is

\begin{eqnarray}
\label{sint}
S_{int}&=&\frac{1}{g^2}\int d^5 z\Big[-\frac{i}{4}D^{\gamma}D^\alpha
A_{\gamma}*
[A^\beta ,D_\beta A_\alpha ]-\frac{1}{12}D^{\gamma}D^\alpha A_{\gamma}*
[A^\beta ,\{A_\beta ,A_\alpha \}]-\nonumber\\&-&
\frac{1}{8}[A^{\gamma},D_{\gamma}A^\alpha ]*[A^\beta ,D_\beta A_\alpha
]+\frac{i}{12}
[A^{\gamma},D_{\gamma}A^\alpha ]*[A^\beta ,\{A_\beta ,A_\alpha \}]+
\nonumber\\&+&\frac{1}{72}
[A^{\gamma},\{A_{\gamma},A^\alpha \}]*[A^\beta ,\{A_\beta ,A_\alpha \}]
\Big]\,.
\end{eqnarray}

The action of the associated Faddeev-Popov ghosts reads

\begin{eqnarray}
\label{sfp}
S_{FP}=\frac{1}{2g^2}\int d^5 z (c'D^\alpha D_\alpha c+ic'*
D^\alpha [A_\alpha ,c])\,,
\end{eqnarray}

\noindent
implying in the propagator

\begin{eqnarray}
\label{pr2}
<c'(z_1)c(z_2)>=-ig^2\frac{D^2}{\Box}\delta^5(z_1-z_2)\,.
\end{eqnarray}

\noindent
We assume that the ghosts are in the adjoint representation.
The total action is, then, given by

\begin{eqnarray}
\label{st}
S_{total}=S+S_{GF}+S_{FP}\,.
\end{eqnarray}

To study the divergence structure of the model we shall start by
determining the superficial degree of divergence $\omega$ associated to
a generic supergraph. Explicitly, $\omega$ receives contributions from the
propagators and, implicitly, from the supercovariant derivatives.
This last dependence can be unveiled by the use of the conversion rule

\begin{eqnarray}
\label{idsg}
D_\alpha D_\beta =i\pa_{\alpha\beta}-C_{\alpha\beta}D^2
\end{eqnarray}

\noindent
and the identity $(D^2)^2=\Box$.  Each loop contributes two power of momentum
to $\omega$. To see how this come about, notice that each integration over $d^3k$ is decreased by one power of momentum when contracting the corresponding loop into a point. It can be seen that, if
$V_3,V_2,V_1,$ and $V_0$ are, respectively, the number of {\it pure} gauge
vertices with three, two, one and none spinor derivatives, then, they
altogether will contribute $\frac{3}{2}V_3+V_2+\hf V_1$.
Furthermore, $V_c$ gauge-ghost vertices will increase $\omega$ by $\hf V_c$.
Each gauge propagator (let their number be $P_A$) lowers $\omega$ by two, each ghost
propagator (let their number be $P_c$) lowers $\omega$ by one. Moving a
supercovariant derivative to an external field decreases $\omega$ by
$\hf$ (let  $N_D$ be the number of spinor derivatives moved to the
external fields).  Putting everything together we may conclude that
$\omega$ is given by

\begin{eqnarray}
\label{o1}
\omega=2L+\frac{3}{2}V_3+V_2+\hf(V_1+V_c)-2P_A-P_c-\hf N_D\,.
\end{eqnarray}

\noindent
The number of the ghost vertices is equal to the number of
the ghost propagators, $P_c=V_c$, since the ghost propagators only form
closed loops. Thus, after using the topological identity $L+V-P=1$ with $P=P_A+P_c$
and $V=V_c+V_0+V_1+V_2+V_3$, we obtain

\begin{eqnarray}
\label{o}
\omega=2-\hf V_c-2V_0-\frac{3}{2}V_1-V_2-\hf V_3-\hf N_D\,.
\end{eqnarray}

\begin{figure}[ht]
\includegraphics{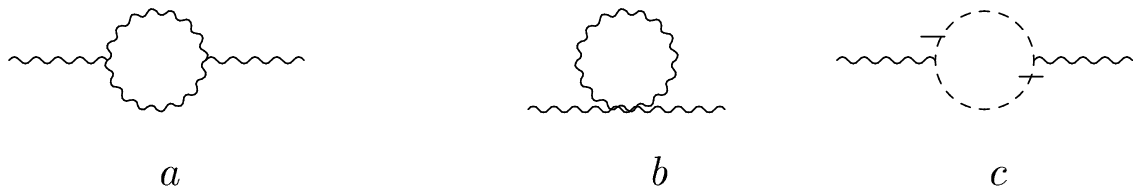}
\caption{Superficially linearly divergent diagrams contributing to the
two-point function of the gauge field.}
\label{Fig1}
\end{figure}

\noindent
This power counting relationship characterizes noncommutative supersymmetric QED$_3$ as an UV super-renormalizable theory.
It is easy to realize that linear divergences may come only from
the graphs with  $V_3=2$, or $V_2=1$,  or $V_c=2$.
These graphs are depicted in  Fig. \ref{Fig1}, they contribute to the
two-point functions of $A^{\a}$ field.
In these graphs, a crossed line corresponds to a factor $D_{\alpha}$
acting on the ghost propagator. A trigonometric factor $e^{ik\wedge
l}-e^{il\wedge k}=2i\sin(k\wedge l)$ originates from each commutator. By denoting the contributions of the graphs in Fig. \ref{Fig1} by $I_{1a}$, $I_{1b}$,  and $I_{1c}$, respectively, we have

\begin{subequations}
\label{1}
\begin{eqnarray}
I_{1a}&=&\frac{1}{32}\int \frac{d^3p}{(2\pi)^3}d^2 \theta_1 d^2\theta_2
\int\frac{d^3k}{(2\pi)^3}\frac{\sin^2(k\wedge p)}{k^2(p-k)^2}
A^\beta (-p,\theta_1)A^{\beta^\prime}(p,\theta_2)\times
\nonumber\\&\times&
\Big[-D^{\g}D^{\a}(C_{\g\g'}\frac{\xi+1}{k^2}+k_{\g\g'}\frac{\xi-1}{k^4}D^2)
D^{\a'}D^{\g'}\delta_{12}\nonumber\\&\times&
D_{\b}(C_{\a\a'}\frac{\xi+1}{k^2}+(p-k)_{\a\a'}\frac{\xi-1}{(p-k)^4}D^2)D_{\b'}
\delta_{12}
\nonumber\\&+&
D^{\g}D^{\a}(C_{\g\a'}\frac{\xi+1}{k^2}+k_{\g\a'}\frac{\xi-1}{k^4}D^2)
D_{\b'}\delta_{12}
\nonumber\\&\times&
D_{\b}(C_{\a\g'}\frac{\xi+1}{k^2}+(p-k)_{\a\g'}\frac{\xi-1}{(p-k)^4}D^2)
D^{\a'}D^{\g'}
\delta_{12}
\Big]\,+\,\cdots\,,
\label{mlett:a1}
\\[10pt]
I_{1b}&=&\frac{1}{3}(\xi+1)\int \frac{d^3p}{(2\pi)^3}d^2 \theta_1
\int\frac{d^3k}{(2\pi)^3}\frac{\sin^2(k\wedge p)}{k^2}
\nonumber\\&\times&\Big[
A^\beta (-p,\theta_1)A_\beta (p,\theta_1)
C_{\gamma\alpha}D^{\gamma}D^\alpha \delta_{12}|_{\theta_1=\theta_2}
-2A^\beta (-p,\theta_1)A_\alpha (p,\theta_1)
C_{\gamma\beta}D^{\gamma}D^\alpha \delta_{12}|_{\theta_1=\theta_2}
\Big]
\nonumber\\&+&\frac{1}{3}(\xi-1)\int \frac{d^3p}{(2\pi)^3}d^2 \theta_1
\int\frac{d^3k}{(2\pi)^3}\frac{\sin^2(k\wedge p)}{k^4}
\nonumber\\&\times&\Big[
A^\beta (-p,\theta_1)A_\beta (p,\theta_1)
k_{\gamma\alpha}D^{\gamma}D^\alpha D^2\delta^2_{12}|_{\theta_1=\theta_2}
-2A^\beta (-p,\theta_1)A_\alpha (p,\theta_1)
k_{\gamma\beta}D^{\gamma}D^\alpha D^2\delta_{12}|_{\theta_1=\theta_2}
\Big]
\nonumber\\
&-&\frac{1}{4}(\xi+1)\int \frac{d^3p}{(2\pi)^3}d^2 \theta_1
\int\frac{d^3k}{(2\pi)^3}\frac{\sin^2(k\wedge p)}{k^2}
\nonumber\\&\times&
A^{\gamma}(-p,\theta_1)A^\beta (p,\theta_1)
\delta^{\alpha}_{\alpha} D_{\gamma 1}D_{\beta 2}
\delta_{12}|_{\theta_1=\theta_2}
\nonumber\\&-&
\frac{1}{4}(\xi-1)\int \frac{d^3p}{(2\pi)^3}d^2 \theta_1
\int\frac{d^3k}{(2\pi)^3}\frac{\sin^2(k\wedge p)}{k^2}
\nonumber\\&\times&
A^{\gamma}(-p,\theta_1)A^\beta (p,\theta_1)
k^{\alpha}_{\alpha} D_{\gamma 1}D^2D_{\beta 2}
\delta_{12}|_{\theta_1=\theta_2}\,+\,\cdots\,,
\label{mlett:b1}\\[10pt]
I_{1c}&=&\frac{1}{2}\int \frac{d^3p}{(2\pi)^3}d^2 \theta_1 d^2\theta_2
\int\frac{d^3k}{(2\pi)^3}\frac{\sin^2(k\wedge p)}{k^2(k+p)^2}
A_\alpha (-p,\theta_1)A_\beta (p,\theta_2)
D^\alpha _1D^2\delta_{12}D^2D^\beta _2\delta_{12}\,.
\label{mlett:c1}
\end{eqnarray}
\end{subequations}

\noindent
Where not otherwise indicated it must be understood that
the supercovariant derivatives act on
the Grassmann variable $\q_1$, also $\delta_{12}=\delta^2(\q_1-\q_2)$. In the expressions for the $I_1$'s the
terms where covariant derivatives act on external fields were omitted
because they do not produce linear divergences and UV/IR mixing
(as we shall shortly verify, such terms give only finite contributions).
In the formulae above they are indicated by the ellipsis.
After some D-algebra transformations we arrive at

\begin{subequations}
\label{contrib1}
\begin{eqnarray}
I_{1a}&=&-\frac{1}{2}\xi\int \frac{d^3p}{(2\pi)^3}d^2 \theta_1
\int\frac{d^3k}{(2\pi)^3}\frac{\sin^2(k\wedge p)}{k^2}
A^\beta (-p,\theta_1)A_\beta (p,\theta_1)
\,+\,\cdots\,,\label{mlett:acontrib1}\\
I_{1b}&=&\frac{1}{2}(1+\xi)\int \frac{d^3p}{(2\pi)^3}d^2 \theta_1
\int\frac{d^3k}{(2\pi)^3}\frac{\sin^2(k\wedge p)}{k^2}
A^\beta (-p,\theta_1)A_\beta (p,\theta_1)\,+\,\cdots\,,\label{mlett:bcontrib1}\\
I_{1c}&=&-\frac{1}{2}\int \frac{d^3p}{(2\pi)^3}d^2 \theta_1
\int\frac{d^3k}{(2\pi)^3}\frac{\sin^2(k\wedge p)}{k^2}
A^\beta (-p,\theta_1)A_\beta (p,\theta_1)\,+\,\cdots\,.\label{mlett:ccontrib1}
\end{eqnarray}
\end{subequations}

\noindent
Hence, the total one-loop two-point function of the gauge superfield,
given by $I_1=I_{1a}+I_{1b}+I_{1c}$, is free from both UV and UV/IR
infrared singularities.  The same situation takes place in the
four-dimensional noncommutative supersymmetric QED
\cite{Bichl,ourqed}.  It is also easy to show that the logarithmically
divergent parts of $I_{1a}$, $I_{1b}$ and $I_{1c}$, which involve
derivatives of the gauge fields, turn out to be proportional to the
integral

\begin{eqnarray}
\label{intl}
\int\frac{d^3k}{(2\pi)^3}\frac{k_{\alpha\beta}\sin^2(k\wedge p)}{k^2(k+p)^2}
\end{eqnarray}

\noindent
and are therefore finite by symmetric integration. Thus, the logarithmic
divergences in $I_{1a},I_{1b}$, and $I_{1c}$ are also absent, i.e., the two-point function of $A^{\a}$ field is {\em finite} in the one-loop approximation. We already mentioned
that linear divergences are possible only for $V_2=1$, or $V_3=2$, or
$V_c=2$. Nevertheless, it is easy to see that two- and
higher-loop graphs satisfying these conditions are just vacuum ones.
Then, there are no linear UV and UV/IR infrared divergences
beyond one-loop and, as consequence, {\em the Green functions are free of nonintegrable infrared divergences at any loop order}.

\begin{figure}[ht]
\includegraphics{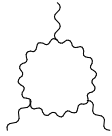}
\caption{A typical logarithmically divergent diagram.}
\label{Fig2}
\end{figure}

We examine next the structure of potentially logarithmic divergent diagrams.
They correspond to $0\leq\omega<1$, which is possible if $V_0=1$, or $V_1=1$, or
$V_2=2$, or $V_c=3,4$, or $V_c=2$ with $V_2=1$, or $V_3=2$ with
$V_2=1$, or $V_3=2$ with $V_c=2$, or $V_3=3,4$, or $V_2=V_3=1$. Notwithstanding, the contributions of these graphs turn out to be very similar among themselves so that the same mechanism of cancellation of divergences applies. As a prototype of this mechanism let us
consider the supergraph with $V_3=3$ in Fig. \ref{Fig2}. Its amplitude in the Feynman gauge reads

\begin{eqnarray}
I_2 & \,=\, &-\frac{1}{3}(\frac{i}{2})^3\int \frac{d^3p_1d^3p_2}{(2\pi)^6}\int
d^2 \theta_1 d^2\theta_2 d^2\theta_3
\nonumber\\&\times&
\int\frac{d^3k}{(2\pi)^3}
\frac{\sin(k\wedge p_1)\sin[k\wedge (p_1+p_2)]\sin[(k+p_1)\wedge p_2]}
{k^2(k+p_1)^2(k+p_1+p_2)^3}
\nonumber\\&\times&
A_\beta (p_1,\theta_1)A_{\beta^\prime}(p_2,\theta_2)A_{\beta^{''}}(-p_1-p_2,\theta_3)
\nonumber\\&\times&
D^{\gamma}D^\alpha D^{\beta^\prime}
\delta_{12}D^{\gamma^{\prime}}D^{\alpha^{\prime}}
D^{\beta^{\prime\prime}}\delta_{23}
D^{\gamma^{\prime\prime}}D^{\alpha^{\prime\prime}}D^{\beta}
\delta_{12}
C_{\gamma\alpha^{\prime}}C_{\gamma^\prime\alpha^{\prime\prime}}
C_{\gamma^{\prime\prime}\alpha}\,.
\end{eqnarray}

\noindent
By using  the relationship (\ref{idsg}) and the identity $\{D_\alpha ,D^2\}=0$ we find that $I_2$  vanishes. The fact that this graph is finite is actually a gauge independent statement. Indeed, in an arbitrary gauge and after D-algebra transformations, $I_2^{(\xi)}$ is given by

\begin{eqnarray}
I_2^{(\xi)} &\,=\, &I_2\,-i\frac{1}{6}\int d^2 \q\int \frac{d^3p_1d^3p_2}{(2\pi)^6}
\int\frac{d^3k}{(2\pi)^3}
\left[\sin(p_1\wedge p_2)-\sum_{i=1}^3\sin(2k\wedge p_i+p_1\wedge p_2)\right]\nonumber\\&\times&
\frac{1}
{k^2(k+p_1)^2(k+p_1+p_2)^3}
k^2\xi(\xi^2-1)
A_\beta (p_1,\q)A_{\beta^\prime}(p_2,\q)\nonumber\\&\times&\left[ k^{\beta^\prime\beta^{''}}
D^\beta A_{\beta^{''}}(p_3,\q)+k^{\beta\beta^{''}}
D^{\beta^\prime}A_{\beta^{''}}(p_3,\q)+k^{\beta^\prime\beta}
D^{\beta^{''}}
A_{\beta^{''}}(p_3,\q)
\right]\,,
\label{other}
\end{eqnarray}

\noindent
whose planar part is proportional to that of the integral in Eq. (\ref{intl}), which is finite. The nonplanar part of $I_2^{(\xi)}$ is composed of two terms, one proportional to

\begin{eqnarray}
\label{intl1}
\int\frac{d^3k}{(2\pi)^3}\frac{k_{\alpha\beta}\cos(2k\wedge p)}{k^2(k+p)^2}\,,
\end{eqnarray}

\noindent
which is evidently finite, and the other proportional to a linear combination of integrals of the form

\bea
\label{intl2}
\int\frac{d^3k}{(2\pi)^3}\frac{k_{\alpha\beta}\sin(2k\wedge p)}{k^4}\,=\,
-\frac{i}{4\pi}\frac{\tilde{p}_{\a\b}}{\sqrt{\tilde{p}^2}}\,.
\eea

\noindent
Here, $\tilde{p}_{\a\b}=\Theta_{mn}p^n(\s^m)_{\a\b}$, and
$\Theta_{mn}$ is the constant antisymmetric matrix characterizing the
noncommutativity of the underlying space-time. As $\Theta^{0i}\,=\,0$,
this last expression does not produce logarithmic divergences, which
confirms the finiteness of the contribution $I_2^{(\xi)}$.

\begin{figure}[ht]
\includegraphics{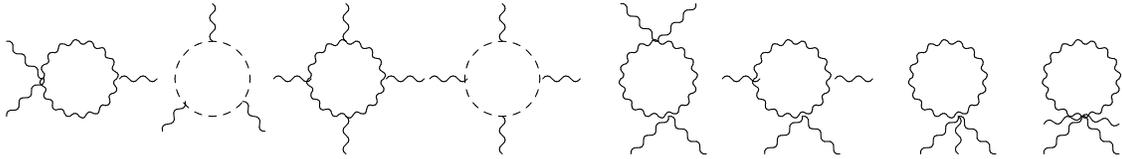}
\caption{Other superficially divergent contributions.}
\label{Fig3}
\end{figure}

The above mechanism also enforces the vanishing of UV logarithmic divergences and of UV/IR infrared logarithmic singularities from the graphs in Fig. \ref{Fig3}. The UV finiteness of all these one-loop graphs may be  proved in an analogous way. For example, in the  Feynman gauge the one-loop graph with $V_2=2$ contains four spinor derivatives and its UV leading contribution is proportional to the finite integral in Eq. (\ref{intl}). A similar situation arises for the one-loop graph with $V_2=V_3=1$. The one-loop graph with $V_3=2$ and $V_2=1$ contains 8 D-factors and, after using the identity $(D^2)^2=\Box$, either a finite contribution proportional to that in Eq. (\ref{intl}) or a finite term in which some derivatives are moved to the external fields could emerge. The others potentially divergent one-loop graphs correspond to $V_c=4$ or $V_3=4$ and for them the same mechanism applies and, hence, they are finite. As it can be checked, the same happens in an arbitrary covariant gauge. The vanishing of UV/IR infrared singularities for all these graphs has the same origin as that for the graph in Fig. \ref{Fig2}.

Up to this point, the net result of our study is that {\em the theory
without matter turns out to be one-loop UV and IR finite}. It is
interesting to note that, in the framework of the background field
method \cite{SGRS,RR}, all contributions to the effective action are
superficially finite. From a formal viewpoint this is caused by the  presence
of two spinor derivatives in the expression for the strength $W_{\a}$ in Eq. (\ref{sstr}),
which makes $N_D\geq 4$ in Eq. (\ref{o}), since loop
corrections must be at least of second order in the background strengths
(compare with \cite{Zanon}). We also remark that Eq. (\ref{o}) implies in the
absence of divergences at three- and higher-loop orders, in agreement with the super-renormalizability of the theory. This concludes our analysis of the ${\cal N}=1$ supersymmetry.

We next study the interaction of the spinor gauge field with matter. To this end we add to (\ref{stot}) the matter action

\begin{eqnarray}
S_m&=&\int d^5 z \left[\hf(D^{\alpha} \bar{\phi}_a+i[\bar{\phi}_a,A^{\alpha}])*
(D_{\alpha} \phi_a-i[A_{\alpha} ,\phi_a])+m\bar{\phi}_a\phi_a
\right]\,.
\end{eqnarray}

\noindent
Here, $\phi_a, a=1,\ldots,N$, are scalar superfields and $\bar{\phi}_a$ their corresponding conjugate ones. We may also write

\begin{eqnarray}
\label{acmat}
S_m&=&\int d^5 z
\Big[-\bar{\phi}_a(D^2-m)\phi_a+i\hf ([\bar{\phi}_a,A^\alpha ]*D_\alpha \phi_a-
D_\alpha \bar{\phi}_a*[A^\alpha ,\phi_a])+\nonumber\\&+&
\hf [\bar{\phi}_a, A^\alpha] *[A_\alpha,\phi_a]\Big]\,.
\end{eqnarray}

\noindent
The free propagator of the scalar fields is

\begin{eqnarray}
<\bar{\phi}_a(z_1)\phi_b(z_2)>=
-i\delta_{ab}\frac{D^2+m}{\Box-m^2}\delta^5(z_1-z_2)\,,
\end{eqnarray}

\noindent
which, in momentum space, reads

\begin{eqnarray}
<\bar{\phi}_a(-k,\theta_1)\phi_b(k,\theta_2)>=i\delta_{ab}\frac{D^2+m}{k^2+m^2}
\delta_{12}\,.
\end{eqnarray}

The superficial degree of divergence when matter is present is given by

\begin{eqnarray}
\label{om}
\omega=2-\hf V_c-2V_0-\frac{3}{2}V_1-V_2-\hf V_3
-\hf E_{\phi}-\hf
V^D_{\phi}-\hf N_D-V^0_{\phi}\,,
\end{eqnarray}

\noindent
where, as before, $V_i$ is the number of pure gauge vertices with $i$
spinor derivatives,
$E_{\phi}$ is the number of external scalar lines,
$N_D$ is the number
of spinor derivatives associated to external lines, $V^D_{\phi}$ is the
number of triple vertices $A^\alpha *\bar \phi_a*{\overleftrightarrow D}_\alpha \phi_a$, and
$V^0_{\phi}$ is the
number of quartic vertices $\phi_a*\bar{\phi}_a*A^\alpha* A_\alpha$.

\begin{figure}[ht]
\includegraphics{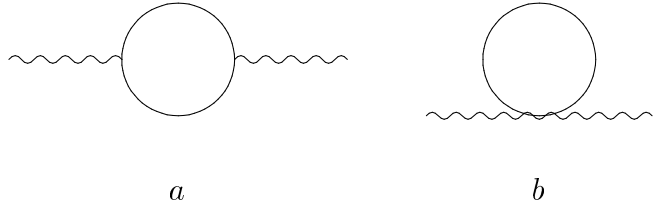}
\caption{One-loop corrections to the self-energy of the spinor gauge field.}
\label{Fig4}
\end{figure}

Graphs can now be split into those with $E_{\phi}=0$ and those with $E_{\phi}\neq0$. The leading UV divergence for those with $E_{\phi}=0$ is $\omega = 3/2$, corresponding to a tadpole graph which vanishes identically. What comes next are graphs with two external $A_{\alpha}$ legs which are UV linearly divergent. They are depicted in Fig. \ref{Fig4}. Graphs with three and four external $A_{\alpha}$ legs are UV logarithmically divergent. The remaining ones are finite. As for the graphs with $E_{\phi}\neq0$, only those with $E_{\phi}=2$ are potentially UV logarithmically divergent, those with $E_{\phi}>2$ are finite.

Graphs with $E_{\phi}=0$ verify the conditions $V^0_{\phi}>0$ or $V^D_{\phi}>0$ which, unless for the tadpole graph already mentioned, imply that $\hf V^D_{\phi}+V^0_{\f}\geq 1$. On the other hand, if $\hf V^D_{\phi}+V^0_{\f}>2$, the corresponding supergraphs are superficially finite, according to (\ref{om}). Since there are no external matter legs, each vertex of the one-loop graph must involve matter. Hence, we arrive at
the following condition for $\omega$ being non negative

\bea
\label{cond1}
1\leq \hf V^D_{\phi}+V^0_{\phi}\leq 2 \,.
\eea

\noindent
The lower limit of the inequality corresponds to $\omega=1$, whereas the upper limit corresponds to $\omega=0$.

The UV linearly divergent case is only realized by the one-loop matter correction to the two-point function of the gauge field $A_{\alpha}$ (Fig. \ref{Fig4}). The graph $a$ in Fig. \ref{Fig4} furnishes,

\begin{eqnarray}
I_{4a} &=&-\int \frac{d^3p}{(2\pi)^3} d^2\theta_1 d^2\theta_2 \int\frac{d^3k}{(2\pi)^3}A^{\a}(-p,
\theta_1)
A^{\b}(p,\theta_2)\sin^2(k\wedge p)\\
&\times&\Big[
D_{\a 1}<\phi_a(1)\bar{\phi}_b(2)>\left(D_{\b 2}<\bar{\phi}_a(1)\f_b(2)>\right)
\nonumber\\
&-&\left(D_{\a 1}D_{\b 2}<\phi_a(1)\bar{\phi}_b(2)>\right)<\bar{\phi}_a(1)\f_b(2)>
\Big]\,,\nonumber
\end{eqnarray}

\noindent
where the indices $1$ and $2$ in the supercovariant derivatives
designate the field to which the $D$ operator is applied.
Taking into account the explicit form of the propagators, we found

\begin{eqnarray}
I_{4a} &=& N \int \frac{d^3p}{(2\pi)^3} d^2\theta_1 d^2\theta_2 \int\frac{d^3k}{(2\pi)^3}A^{\a}
(-p,\theta_1)
A^{\b}(p,\theta_2)\sin^2(k\wedge p)\nonumber\\&\times&\Big[
\frac{D_{\a 1}(D^2_1+m)}{k^2+m^2}\delta_{12}
\frac{(D^2_1+m)D_{\b 2}}{(k+p)^2+m^2}\delta_{12}
\nonumber\\&-&
\frac{D_{\a 1}(D^2_1+m)D_{\b 2}}{k^2+m^2}\delta_{12}
\frac{D^2_1+m}{(k+p)^2+m^2}\delta_{12}
\Big]\,,
\end{eqnarray}

\noindent
which, after using $D_{\b 2}\delta_{12}=-D_{\b 1}\delta_{12}$, can be cast as

\begin{eqnarray}
\label{expr}
I_{4a} &=&N \int \frac{d^3p}{(2\pi)^3} d^2\theta_1 d^2\theta_2 \int\frac{d^3k}{(2\pi)^3}
J(k,p)
\nonumber\\&\times&\Big[
2(D^2_1+m)\delta_{12}
D_{\a 1}(D^2_1+m)D_{\b 1}\delta_{12}
A^{\a}(-p,\theta_1) A^{\b}(p,\theta_2)
+\nonumber\\&+&
(D^2_1+m)\delta_{12} (D^2_1+m)D_{\b 1}\delta_{12}
(D^{\a}A_{\a})(-p,\theta_1) A^{\b}(p,\theta_2)
\Big]\,,
\end{eqnarray}

\noindent
where we have introduced the notation

\begin{eqnarray}
J(k,p)=\frac{\sin^2(k\wedge p)}{(k^2+m^2)\left[(k+p)^2+m^2\right]}\,.
\end{eqnarray}

\noindent
It is convenient to split $I_{4a}$ into two parts, $I_{4a}\,=\,I_{4a}^{(1)}+I_{4a}^{(2)}$, where $I_{4a}^{(1)}$ and $I_{4a}^{(2)}$ are, respectively, associated to the first and second terms in the large brackets in the right hand side of Eq. (\ref{expr}). It is straightforward to verify that

\begin{eqnarray}
&&I_{4a}^{(1)} \,=\, 2N \int \frac{d^3p}{(2\pi)^3} d^2\q  \int\frac{d^3k}{(2\pi)^3} J(k,p)
\nonumber\\
&&\times \Big[-
(k^2+m^2)C_{\a\b} A^{\a}(-p,\q)A^{\b}(p,\q)
+(k_{\a\b}-mC_{\a\b})(D^2A^{\a}(-p,\q)) A^{\b}(p,\q)
\Big]\,.\label{1a}
\end{eqnarray}

\noindent
For the second term in the right hand side of Eq. (\ref{expr}) one analogously finds

\begin{eqnarray}
I_{4a}^{(2)}&=& N \int \frac{d^3p}{(2\pi)^3} d^2 \q \int\frac{d^3k}{(2\pi)^3} J(k,p)
\Big[D^{\gamma}D^{\a}A_{\a}(-p,\q)
(k_{\gamma\b}-mC_{\gamma\b})A^{\b}(p,\q)
\Big]\,.\label{2}
\end{eqnarray}

\noindent
By adding Eqs. (\ref{1a}) and (\ref{2}) we can cast the contribution
from the graph $a$ in Fig. \ref{Fig4} as

\begin{eqnarray}
\label{s1}
I_{4a} &=& 2N \int \frac{d^3p}{(2\pi)^3} d^2\q \int \frac{d^3k}{(2\pi)^3}
\frac{\sin^2(k\wedge p)}{(k^2+m^2)\left[(k+p)^2+m^2\right]}
\nonumber\\&\times&\Big[-
(k^2+m^2)C_{\a\b} A^{\a}(-p,\q)A^{\b}(p,\q)
+(k_{\a\b}-mC_{\a\b})\left[D^2A^{\a}(-p,\q)\right] A^{\b}(p,\q)
\nonumber\\&+&\hf D^{\gamma}D^{\a}A_{\a}
(k_{\gamma\b}-mC_{\gamma\b})A^{\b}(p,\q)
\Big]\,.
\end{eqnarray}

The algebraic manipulations for the graph $b$ in Fig. \ref{Fig4} are simpler and yield

\begin{eqnarray}
\label{s2}
I_{4b}&=&2N\int \frac{d^3p}{(2\pi)^3} d^2\q \int \frac{d^3k}{(2\pi)^3}\frac{\sin^2(k\wedge p)}{(k+p)^2+m^2}
C_{\a\b} A^{\a}(-p,\q)A^{\b}(p,\q)\,.
\end{eqnarray}

The complete correction to the two-point function is, therefore,

\begin{eqnarray}
\label{stot}
I_4 &=& 2N \int \frac{d^3p}{(2\pi)^3} d^2\q \int \frac{d^3k}{(2\pi)^3}
\frac{\sin^2(k\wedge p)}{(k^2+m^2)\left[(k+p)^2+m^2\right]}
\nonumber\\&\times&
(k_{\gamma\beta}-mC_{\gamma\beta})\Big[(D^2A^{\gamma}(-p,\q)) A^{\b}(p,\q)
+\hf D^{\gamma}D^{\a}A_{\a}(-p,\q) A^{\b}(p,\q)
\Big]\,.
\end{eqnarray}

\noindent
We stress that the dangerous linear divergences have disappeared,
i.e., the two-point function of $A^{\a}$ field turns out to be free of
UV/IR infrared singularities and, moreover, finite. This two-point
function can be used for deriving the effective propagators in the
$\frac{1}{N}$ expansion\cite{scpn}.

\begin{figure}[ht]
\includegraphics{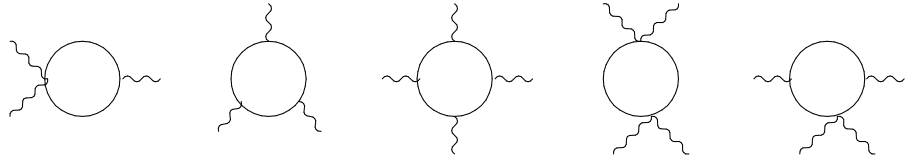}
\caption{Contributions to the three and four point functions of the spinor gauge field.}
\label{Fig6}
\end{figure}

It remains to consider the graphs with $\omega=0$. It follows from
(\ref{cond1}), that the only remaining one-loop logarithmically
divergent graphs involving matter are those ones depicted in
Fig.~\ref{Fig6}. Nevertheless, a direct calculation shows that the
planar contributions of the first three of these supergraphs are
proportional to the integral in Eq. (\ref{intl}) whose divergent part
is known to vanish. The divergent parts of their nonplanar
contributions vanish in a way similar to that of the graphs in
Figs. \ref{Fig2} and \ref{Fig3}. The fourth graph  is evidently finite.
The last graph is finite by the same reason as the first three.

\begin{figure}[ht]
\includegraphics{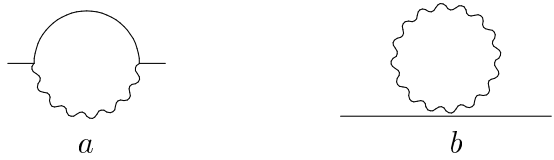}
\caption{One-loop corrections to the self-energy of the $\phi$ field.}
\label{Fig5}
\end{figure}

We shall next deal with the graphs with $E_{\phi}>0$. Such graphs do not contain linear divergences, according to Eq. (\ref{om}). Furthermore, the number of external scalar legs must be even since any vertex carries an even number of scalar
fields, and only an even number of them can be contracted into
propagators. As stated before, the logarithmic divergences in this case are
possible only for $E_{\f}=2, V^D_{\f}=2$ and for
$E_{\f}=2,V^0_{\f}=1$. These graphs are shown in Fig. \ref{Fig5}.
The graph $a$ in \ref{Fig5} gives the contribution

\begin{eqnarray}
I_{6a} &=& 2g^2
\int \frac{d^3p}{(2\pi)^3} d^2\theta_1 d^2\theta_2 \int \frac{d^3k}{(2\pi)^3}
\bar{\phi}_a(-p,\theta_1)\phi_a(p,\theta_2)
\frac{\sin^2(k\wedge p)}{k^2\left[(k+p)^2+m^2\right]}
D^\alpha (D^2-m)D^\beta \delta_{12}\nonumber\\&\times&
\Big[\hf (\xi+1)C_{\alpha\beta}+
\hf(\xi-1)\frac{k_{\alpha\beta}}{k^2}D^2\Big]\delta_{12}+\ldots\,\,.
\end{eqnarray}

\noindent
As before, the ellipsis stands for manifestly finite terms. After some simplifications, one obtains

\begin{eqnarray}
I_{6a}&=&-2\xi g^2 m
\int \frac{d^3p}{(2\pi)^3} d^2\q \int \frac{d^3k}{(2\pi)^3}
\bar{\phi}_a(-p,\q)\phi_a(p,\q)\frac{\sin^2(k\wedge p)}{k^2\left[(k+p)^2+m^2\right]}\,,
\end{eqnarray}

\noindent
which is finite. The second graph in Fig. \ref{Fig5} yields the amplitude

\begin{eqnarray}
I_{6b}\,=\,(\xi-1)
\int \frac{d^3p}{(2\pi)^3} d^2\q_1 \int \frac{d^3k}{(2\pi)^3}
\bar{\phi}_a(-p,\theta_1)\phi_a(p,\theta_2)\frac{k^{\alpha}_{\phantom{a}\alpha}}{k^4}
\sin^2(k\wedge p)
D^2\delta_{12}|_{\theta_1=\theta_2} \,,
\end{eqnarray}

\noindent
which vanishes identically because of ${k^{\alpha}_{\phantom{a}\alpha}}\,=\,0$.

Therefore the two-point function of the scalar field is
free from UV/IR mixing and, moreover, finite in any covariant gauge.
It follows from Eq. (\ref{om}) that the supergraphs with two or more
external scalar legs and one or more gauge legs are also superficially finite.

To sum up we conclude that
{\em the three-dimensional noncommutative supersymmetric QED is
one-loop UV and UV/IR infrared finite both without and with matter}.
A natural development of this work consists in the investigation of the
possibility of appearance of divergences at two-loop order.
Other possible developments are a detailed
study of the  $1/N$ expansion for the model involving many scalar
fields and the analysis of spontaneous symmetry breaking and the Higgs
mechanism.

{\bf Acknowledgements.} This work was partially supported by Funda\c
c\~ao de Amparo \`a Pesquisa do Estado de S\~ao Paulo (FAPESP) and
Conselho Nacional de Desenvolvimento Cient\'\i fico e Tecnol\'ogico
(CNPq). H. O. G. also acknowledges support from PRONEX
under contract CNPq 66.2002/1998-99. A. Yu. P. has been supported by
FAPESP, project No. 00/12671-7.

\newpage

\end{document}